\documentclass[prb,preprint,superscriptaddress,twocolumn,10pt]{revtex4-1}

\usepackage{amsmath} 
\usepackage{amsfonts}
\usepackage{graphicx}
\usepackage{esvect}
\usepackage{mathtools}
\usepackage{subcaption}

\usepackage{xcolor}
\pagecolor{white}

\usepackage{appendix}

\usepackage{siunitx} 
\usepackage{placeins}

\usepackage{float}

\usepackage{pbox}

\begin{document}

\title{Elementary Methods for \\ Infinite Resistive Networks with Complex Topologies}

\author{Tung X. Tran}
\affiliation{Hanoi-Amsterdam High School, 01 Hoang Minh Giam Str., Trung Hoa Nhan Chinh, Cau Giay Dist., Hanoi 100000, Vietnam}

\author{Linh K. Nguyen}
\affiliation{Lam Son High School, 307 Le Lai Str., Dong Son Dist., Thanh Hoa 440000, Vietnam} 

\author{Quan M. Nguyen}
\affiliation{Hanoi-Amsterdam High School, 01 Hoang Minh Giam Str., Trung Hoa Nhan Chinh, Cau Giay Dist., Hanoi 100000, Vietnam}

\author{Chinh D. Tran}
\affiliation{Hung Vuong High School, 70 Han Thuyen Str., Tan Dan, Viet Tri, Phu Tho 35000, Vietnam} 

\author{Truong H. Cai}
\affiliation{School of Engineering, Brown University, Providence, RI 02912, USA}

\author{Trung V. Phan}
\email{tvphan@princeton.edu}
\affiliation{Department of Physics, Princeton University, Princeton, NJ 08544, USA}

\date{\today}

\begin{abstract}
Finding the equivalent resistance of an infinite ladder circuit is a classical problem in physics. We expand this well-known challenge to new classes of network topologies, in which the unit cells are much more entangled together. The exact analytical results there can still be obtained with elementary methods. These topology classes will add layers of complexity and much more diversity to a very popular kind of physics puzzles for teachers and students. 
\end{abstract}

\maketitle

\section{Introduction}

\begin{figure}[!htb]
\centering
\includegraphics[width=0.45\textwidth]{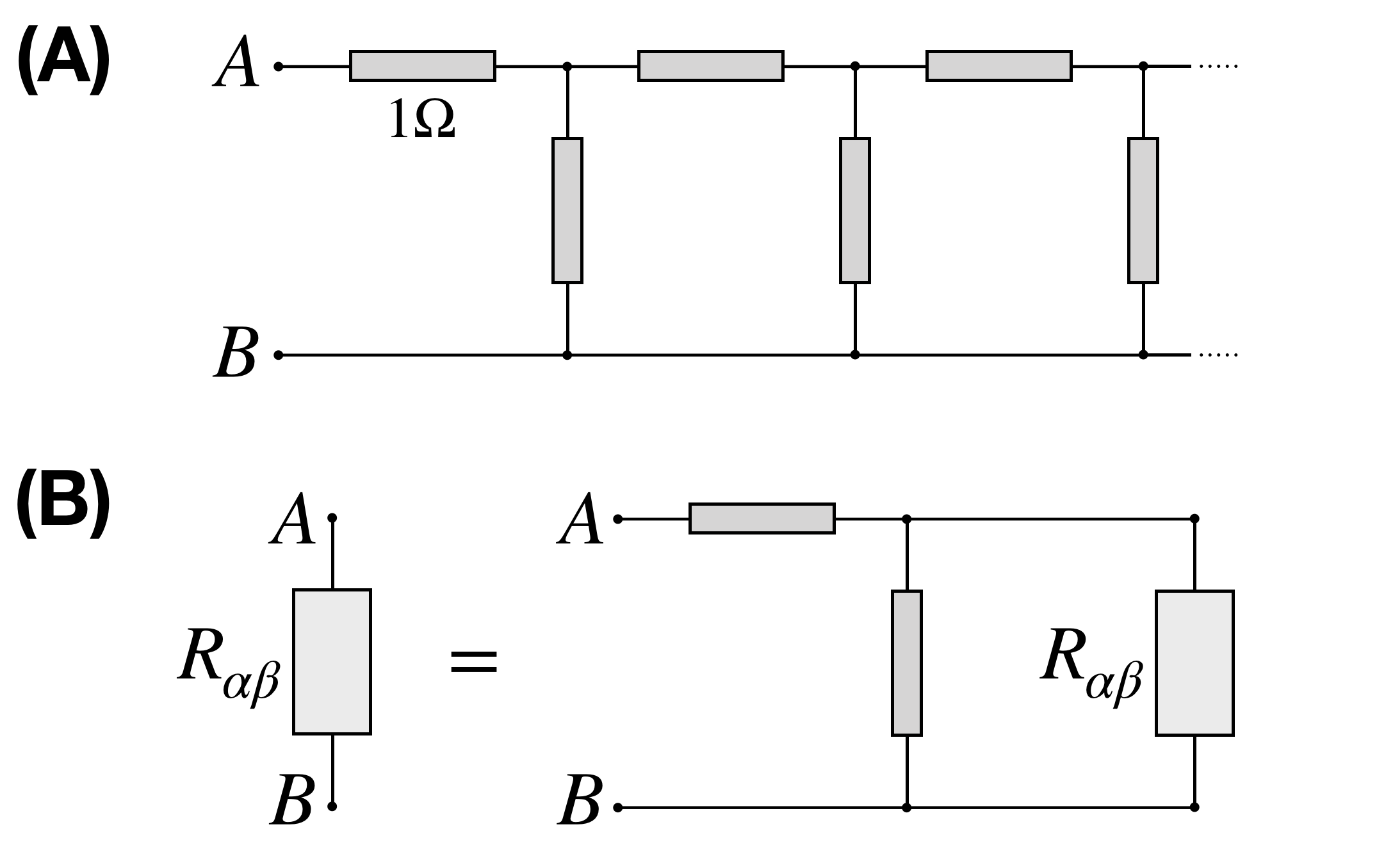}
\caption{(A) The infinite ladder topology network of $1\Omega$ resistors. (B) The trick of adding one more unit cell to the network, where the equal sign is valid under the assumption of convergence.} 
\label{ladder_topology}
\end{figure}

The equivalent resistance $R_{AB}$ of the infinite ladder network composed of identical $1\Omega$ resistors, shown in Fig. \ref{ladder_topology}, can be calculated elegantly by adding one more unit cell (a ladder step) in the front \cite{feynman_lecture}:
\begin{equation}
R_{AB} = R_{\alpha \beta} = 1 + \frac{R_{\alpha \beta}}{1+R_{\alpha \beta}} \ \ . \ \ 
\end{equation}
Here we make the assumption that as the number of ladder steps go to infinity, the equivalent resistance will converge. There are two possible solutions, $R_{AB}=(1\pm \sqrt{5})/2 \Omega$, and by getting rid of the unphysical one with negative value we arrive at the answer to be uniquely $R_{AB}=(1+\sqrt{5})/2 \Omega$ which is equal to the golden ratio.

This problem is addressed in many introductory physics courses in college, and was also introduced to high school students in the very first International Physics Olympiad (Poland 1967). Despite of being widely known for a long time, there aren't many variations: the unit cells are always linearly linked to the one before or after it via $\mathcal{N}=2$ nodes. 

\begin{figure}[!htb]
\centering
\includegraphics[width=0.45\textwidth]{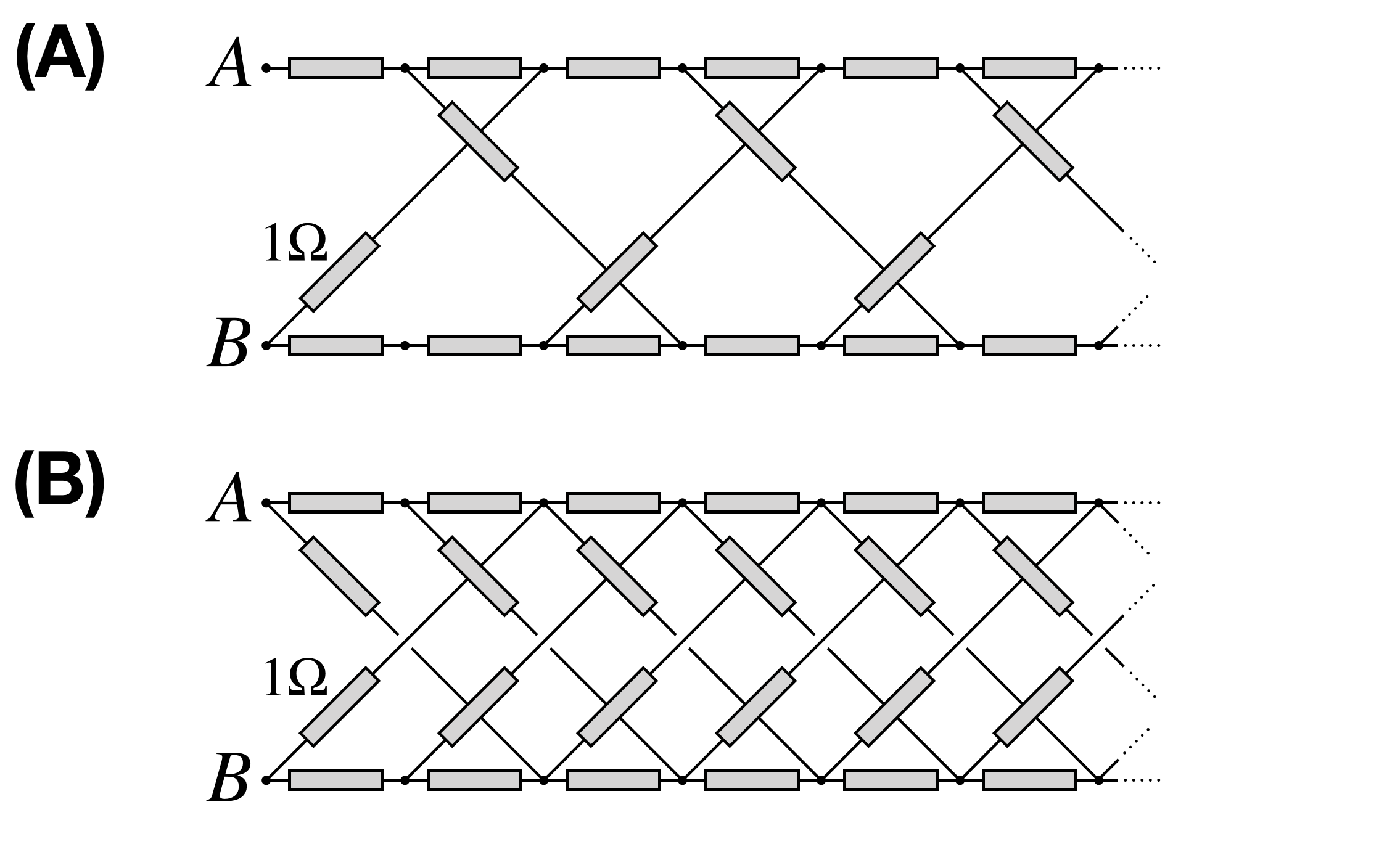}
\caption{(A) The $\mathcal{N}=3$ infinite asymmetric twisted ladder network of $1\Omega$ resistors. (B) The $\mathcal{N}=4$ infinite symmetric twisted ladder network of $1\Omega$ resistors.}
\label{twist_ladder_topologies}
\end{figure}

In this paper we will explore new classes of resistive networks with topologies more entangled, where the linear linking between unit cells has $\mathcal{N}>2$ nodes. To our knowledge this kind of electrical circuits is rarely mentioned, though they can still be solved by elementary methods using resistive relations come from superposition theorem and reciprocal theorem. Starting with estimating the results in Section \ref{rayleigh}, we then develop modified methods for infinite networks and consider two twisted ladder topologies as representatives (see Fig. \ref{twist_ladder_topologies}). Alongside, we will show some examples of network topologies that can be solved similarly, with all results being found analytically and confirmed numerically. We hope to provide teachers and students with more physics puzzles of this kind to explore.

\section{Rough estimations using Rayleigh's Monotonicity Law \label{rayleigh}} 

\begin{figure}[!htb]
\centering
\includegraphics[width=0.45\textwidth]{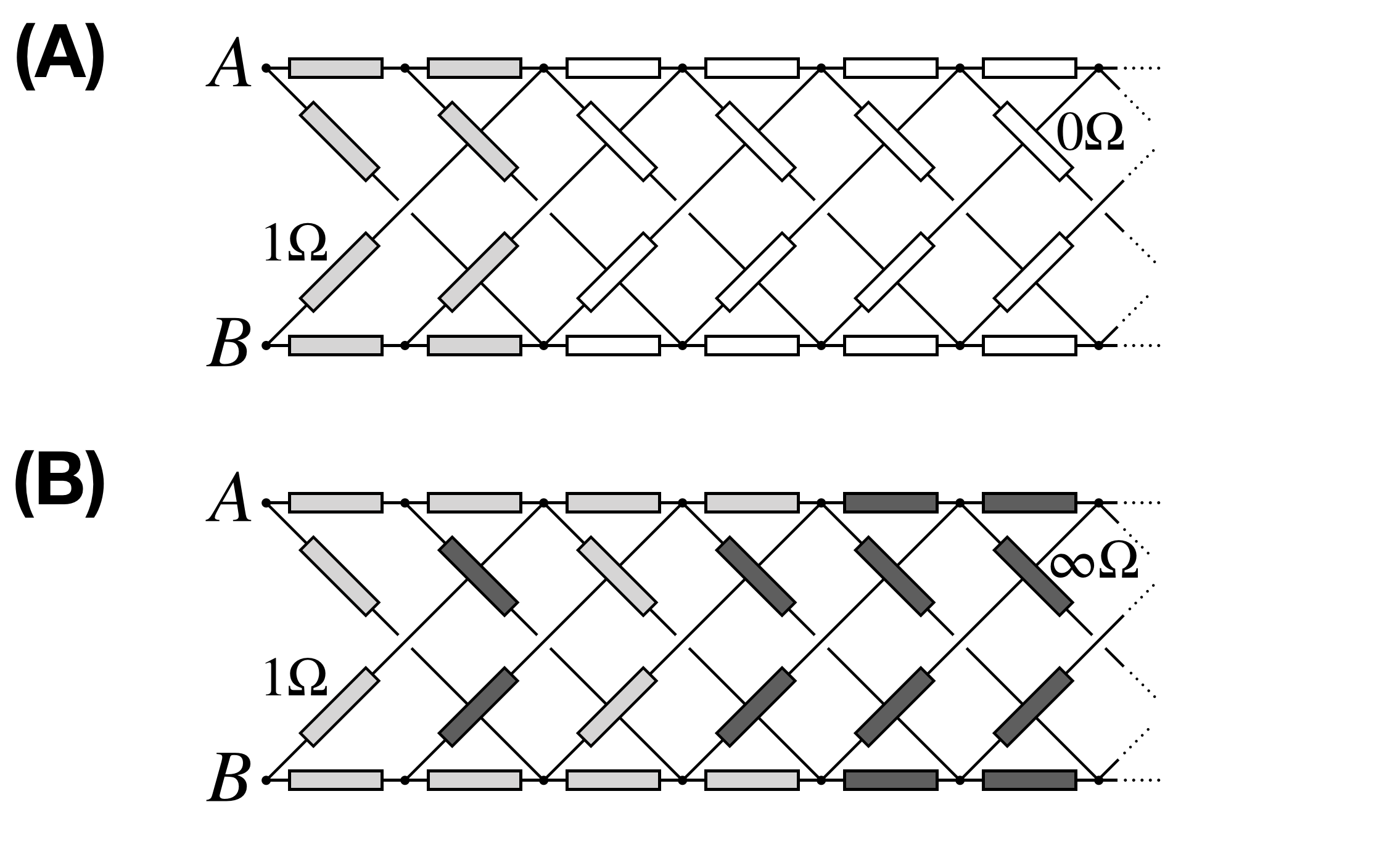}
\caption{Estimations for $R^{(sym)}_{AB}$ of the infinite resistive network with symmetric twisted ladder topology. (A) For the lower bound, we set some resistors to $0\Omega$ and get $R^{(sym)}_{AB} > 6/5 \Omega$. (B) For the upper bound, we set some resistors to $\infty \Omega$ and get $R^{(sym)}_{AB} < 17/12 \Omega$.}
\label{rayleigh_dense}
\end{figure}

Even without arriving at the solutions to these problems, students should be encouraged to make guesses about what the resistances might be. We can make some rough estimations for the equivalent resistance values $R^{(asym)}_{AB}$ and $R^{(sym)}_{AB}$ of those infinite twisted ladders, using Rayleigh's monotonicity law which states that when the resistance in one
part of a circuit increases, the effective resistance also increases \cite{maxwell_treatise}. By setting some resistors to the extremes of $0\Omega$ and $\infty \Omega$ (see Fig. \ref{rayleigh_thin} and Fig. \ref{rayleigh_dense}), we can achieve within the level of $\pm 10\%$ uncertainty:

\begin{figure}[!htb]
\centering
\includegraphics[width=0.45\textwidth]{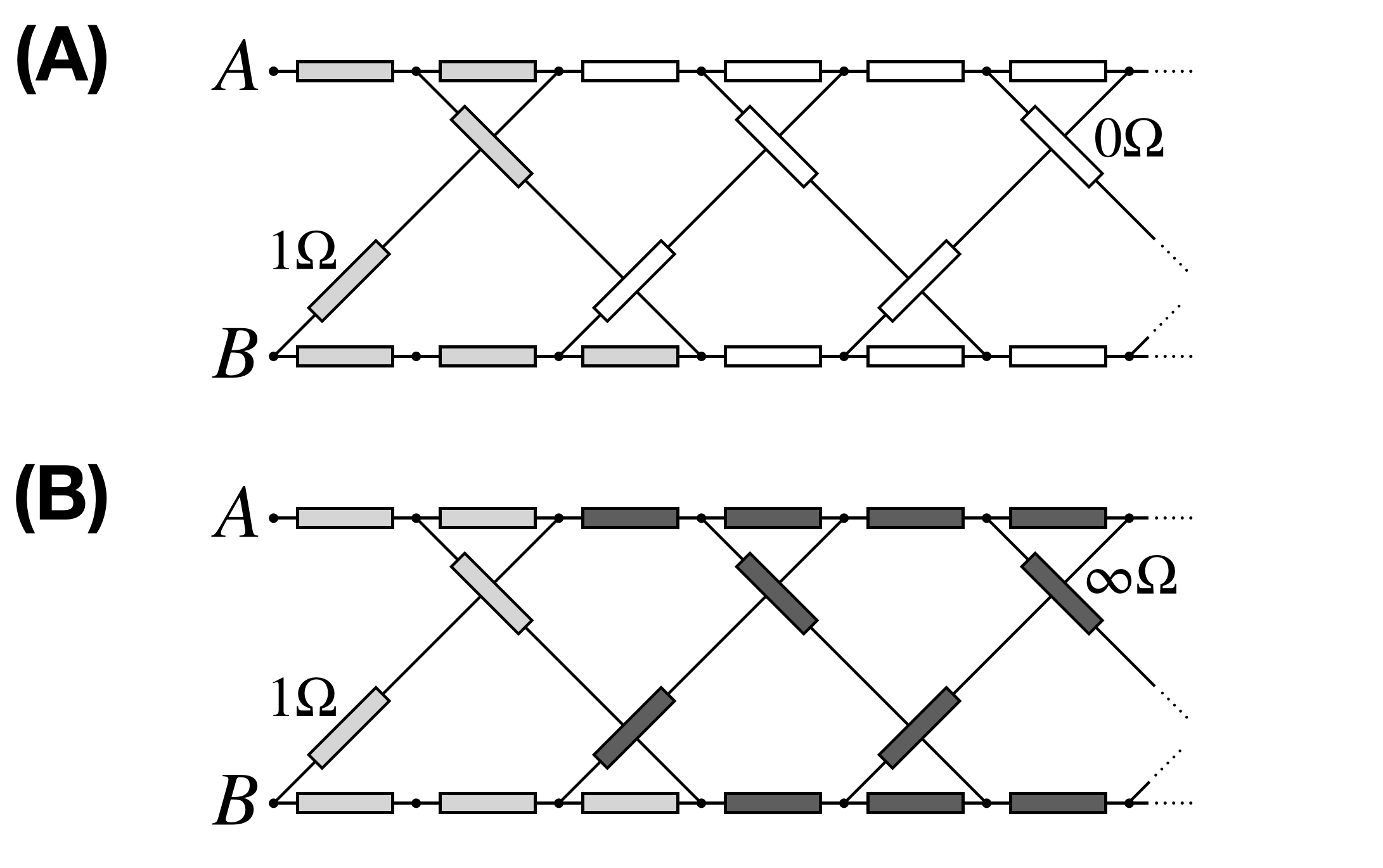}
\caption{Estimations for $R^{(asym)}_{AB}$ of the infinite resistive network with asymmetric twisted ladder topology.(A) For the lower bound, we set some resistors to $0\Omega$ and get $R^{(asym)}_{AB} > 13/6 \Omega$. (B) For the upper bound, we set some resistors to $\infty \Omega$ and get $R^{(asym)}_{AB} < 7/3 \Omega$.}
\label{rayleigh_thin}
\end{figure}

\begin{subequations}
\begin{align}
&7/3 \Omega \approx 2.33 \Omega  > R^{(asym)}_{AB} >  13/6 \Omega \approx 2.17 \Omega  \ \ , \ \ 
\label{estimate_thin}
\\
&17/12 \Omega \approx 1.42 \Omega  >  R^{(sym)}_{AB} >  6/5 \Omega = 1.20 \Omega \ \ . \ \ 
\label{estimate_dense}
\end{align}
\end{subequations}

\FloatBarrier

\section{A Toolkit for Resistive Networks \label{toolkit}} 

\begin{figure}[!htb]
\centering
\includegraphics[width=0.45\textwidth]{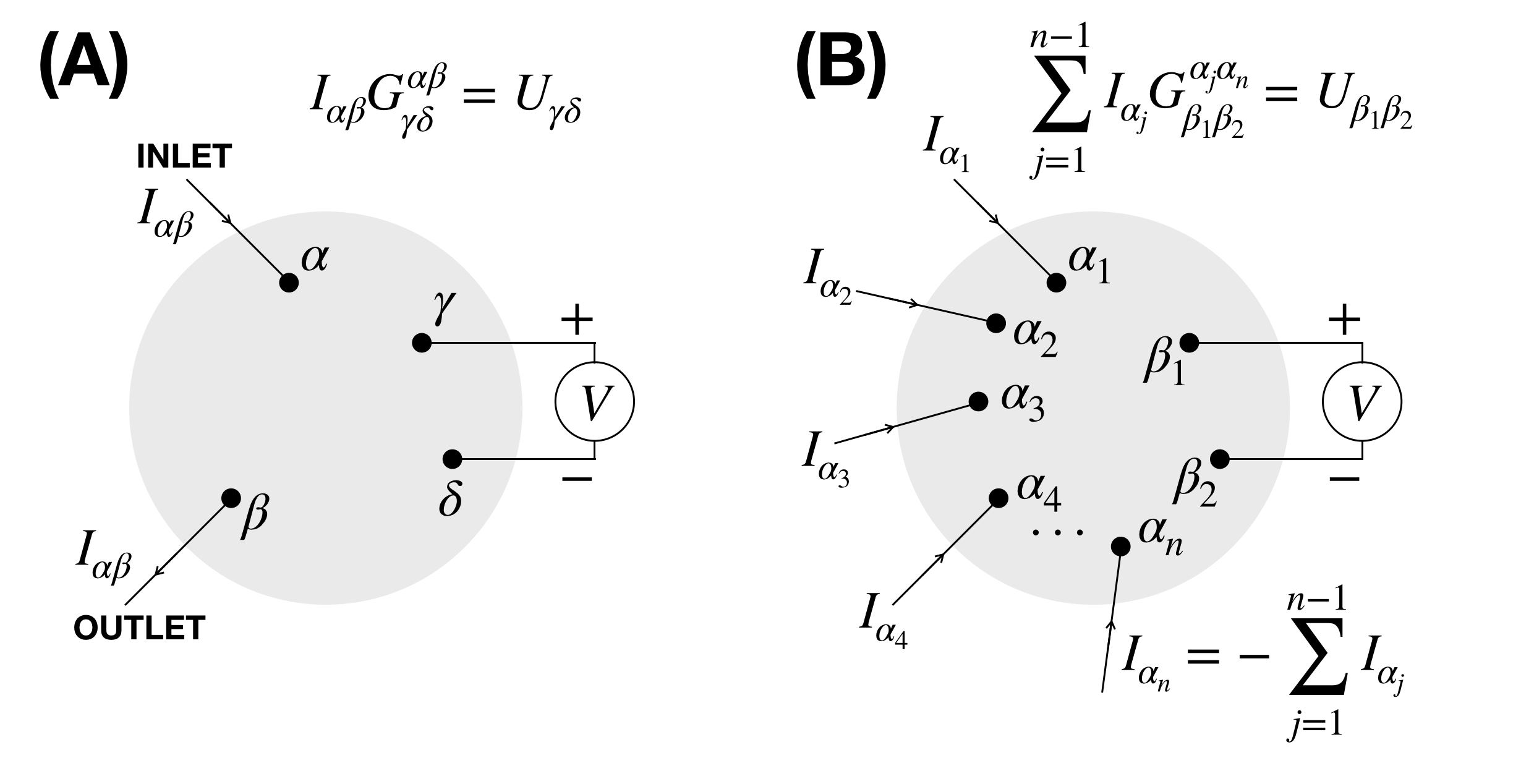}
\caption{(A) The generalized resistance $G^{\gamma \delta}_{\alpha \beta}$ is defined from the input current $I_{\alpha \beta}$ and the induced voltage difference $U_{\gamma \delta}$. It should be noted that the Greek labels $\alpha$, $\beta$, ... are not used as dummy indices. (B) A generalized current-voltage relation.}
\label{generalized_resistance}
\end{figure}

In this section we create a toolkit for resistive networks, using only superposition theorem and reciprocal theorem. Define the \textit{generalized resistance} $G^{\alpha \beta}_{\gamma \delta}$ from the voltage different $U_{\gamma \delta} = V_\gamma - V_\delta$ between nodes $\gamma$ and $\delta$ when there's only a current $I_{\alpha \beta}$ comes into node $\alpha$ and goes out from node $\beta$ (see Fig. \ref{generalized_resistance}A):
\begin{equation}
G^{\alpha \beta}_{\gamma \delta} = U_{\gamma \delta} / I_{\alpha \beta} \ \ . \ \ 
\end{equation}
The resistance $R_{\alpha \beta}$ between nodes $\alpha$ and $\beta$ is related to this as $R_{\alpha \beta} = G^{\alpha \beta}_{\alpha \beta}$. 

Follow from the definition, switching the voltage's nodes and the current's nodes gives the \textit{permutation rule}:
\begin{equation}
G^{\alpha \beta}_{\gamma \delta} = - G^{\alpha \beta}_{\delta \gamma} =  -G^{\beta \alpha}_{\gamma \delta} \ \ . \ \ 
\label{G_switching} 
\end{equation}
For an arbitrary node $\sigma$, using the additivity of voltage difference and superposition theorem for current, we have the \textit{summing rule}:
\begin{equation}
G^{\alpha \beta}_{\gamma \delta} = G^{\alpha \beta}_{\gamma \sigma} + G^{\alpha \beta}_{\sigma \delta}  \ \ , \ \ G^{\alpha \beta}_{\gamma \delta} = G^{\alpha \sigma}_{\gamma \delta} + G^{\sigma \beta}_{\gamma \delta}  \ \ . \ \ 
\label{G_summing}
\end{equation}
With reciprocal theorem \cite{skilling_circuit} (which can be derived from superposition theorem and uniqueness theorem), we arrive at the \textit{flipping rule}:
\begin{equation}
G^{\alpha \beta}_{\gamma \delta} = G^{\gamma \delta}_{\alpha \beta} \ \ . \ \ 
\label{G_reciprocal} 
\end{equation}
Some students might be more familiar with source transformation \cite{nilsson_intro}, which can also be used to obtain the above result and indeed is a direct application of reciprocal theorem.

From equations \eqref{G_summing} and \eqref{G_reciprocal}, every generalized resistance $G^{\alpha \beta}_{\gamma \delta}$ can be written in terms of resistances as:
\begin{equation}
G_{\alpha\beta}^{\gamma\delta} = \frac{R_{\alpha\delta} + R_{\beta\gamma} - R_{\alpha\gamma} - R_{\beta\delta}}2 \ \ . \ \ 
\label{G_to_R}
\end{equation}
For a ``black box'' of resistors with $n$ terminals $\alpha_1$, $\alpha_2$, ... $\alpha_n$, this implies that knowing the values of $n(n-1)/2$ resistances $\{ R_{\alpha_j \alpha_{k \neq j}} \}$ is enough to determine all characteristics $\{ G_{\alpha_j \alpha_k}^{\alpha_l \alpha_m} \}$ of the ``black box'', thus knowing the global property of the network. In other words, internal resistor networks with identical set of $\{ R_{\alpha_j \alpha_k} \}$ are equivalent even though their topologies are different. 

\begin{figure}[!htb]
\centering
\includegraphics[width=0.45\textwidth]{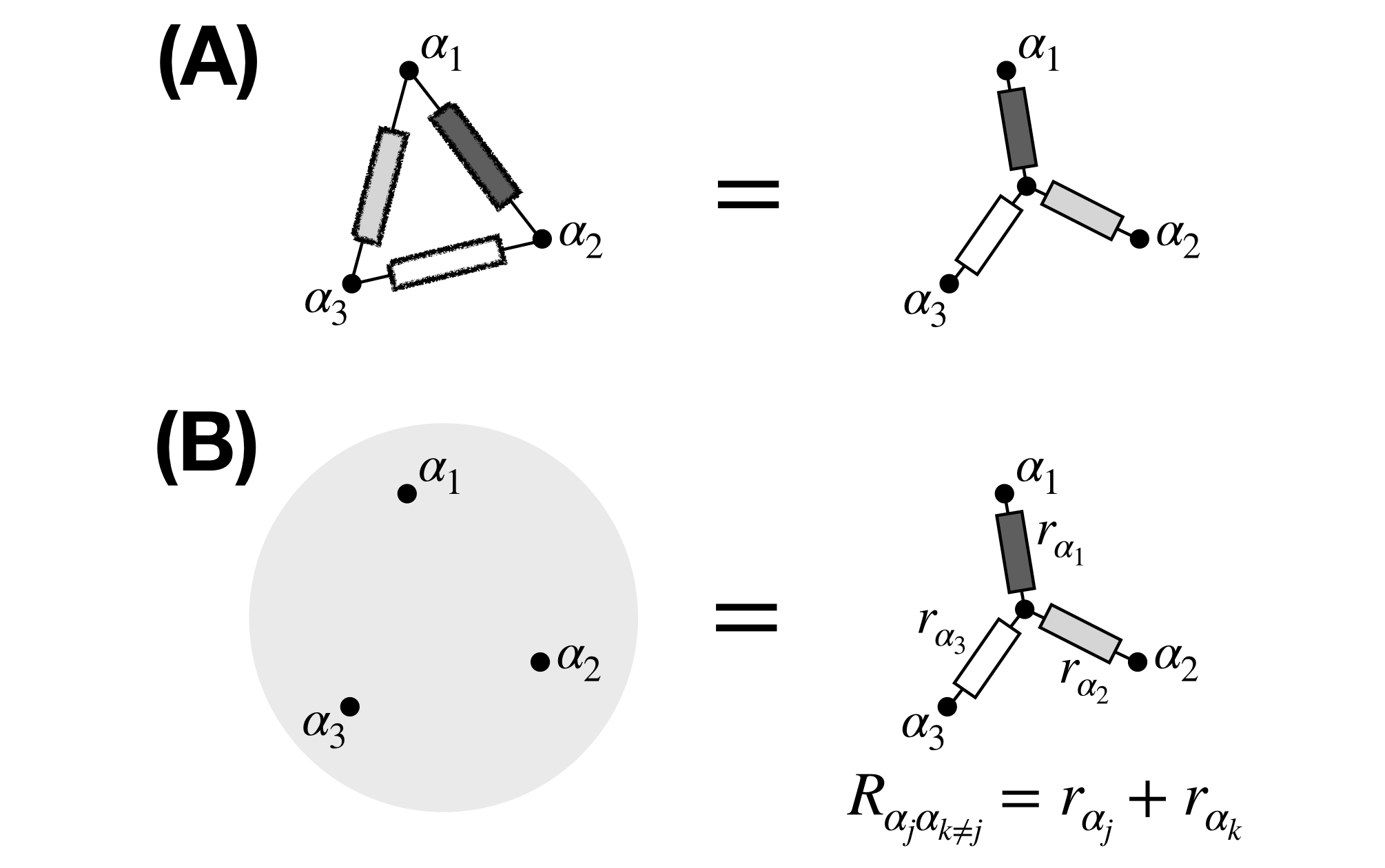}
\caption{(A) The $\Delta$-Y transformation. (B) An extended version of $\Delta$-Y transformation, where $\Delta$ can be generalized to any internal topology.}
\label{wye_delta}
\end{figure}

For $n=3$, $\Delta$-Y are the simplest topologies possible, and it is known that local transformation from $\Delta$ topology to Y topology \cite{Kennelly_YD} (see Fig. \ref{wye_delta}A) does not change any global property. An extended version of $\Delta$-Y transformation exists (see Fig. \ref{wye_delta}B): an arbitrary internal network can be simplified to either $\Delta$ or Y topology, since there always exists a star network that satisfies \eqref{G_to_R} for any $G^{\alpha \beta}_{\gamma \delta}$ with $\{ \alpha, \beta , \gamma, \delta \} \subset \{ \alpha_1, \alpha_2 , \alpha_3 \}$. This simplification, illustrated in Fig \ref{wye_delta}, will be applied directly in Section \ref{section_thin}.

As it will be shown below, this method is easy to be done for $n=3$. When $n>3$, as the simplest topologies of which an arbitrary internal network can be simplified down to are either a mesh or specific combinations of many stars, it becomes highly convoluted to analyze even the simplified topologies in relation to the new unit cell. In this case, a method independent of internal topology becomes preferable.

\section{Calculating the Equivalent Resistances \label{calculate}}
In this section, using the two methods above, we will show that $R_{AB}^{(asym)} = \sqrt{2\sqrt{7}}$ and $R_{AB}^{(sym)} = \sqrt{-1+2\sqrt{2}}$, along with analytical results for other $\mathcal{N}>2$ entangled topologies.

\subsection{$\mathcal{N}=3$ Infinite Asymmetric twisted Ladder \label{section_thin}}

\begin{figure}[!thb]
\centering
\includegraphics[width=0.45\textwidth]{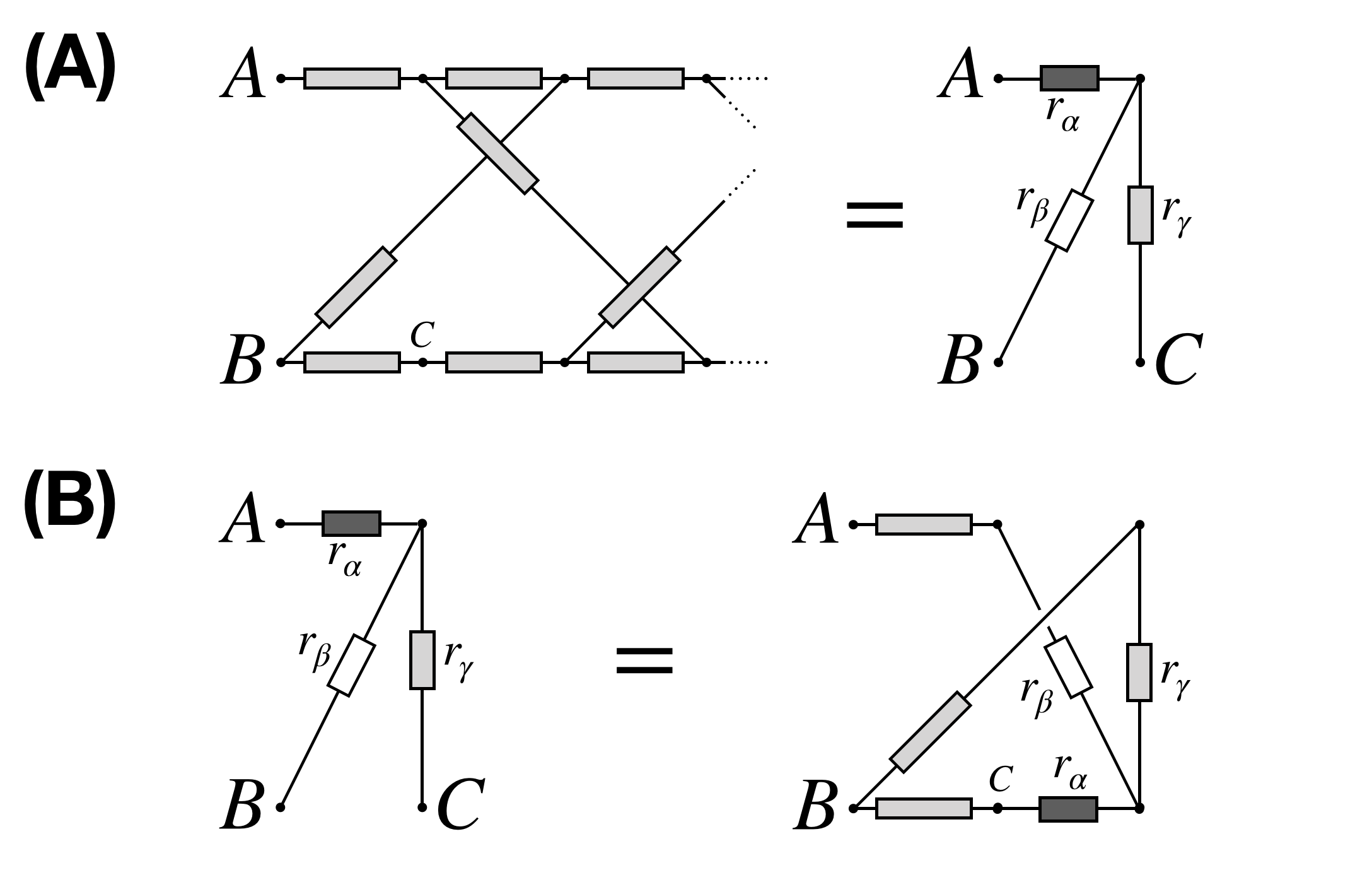}
\caption{(A) Apply the extended version of $\Delta$-Y transformation to the infinite resistive network between nodes A, B and C. (B) The trick of adding one more unit cell to the network, where the equal sign is valid under the assumption of convergence.}
\label{repeat_thin}
\end{figure}

\begin{figure}[!htb]
\centering
\includegraphics[width=0.45\textwidth]{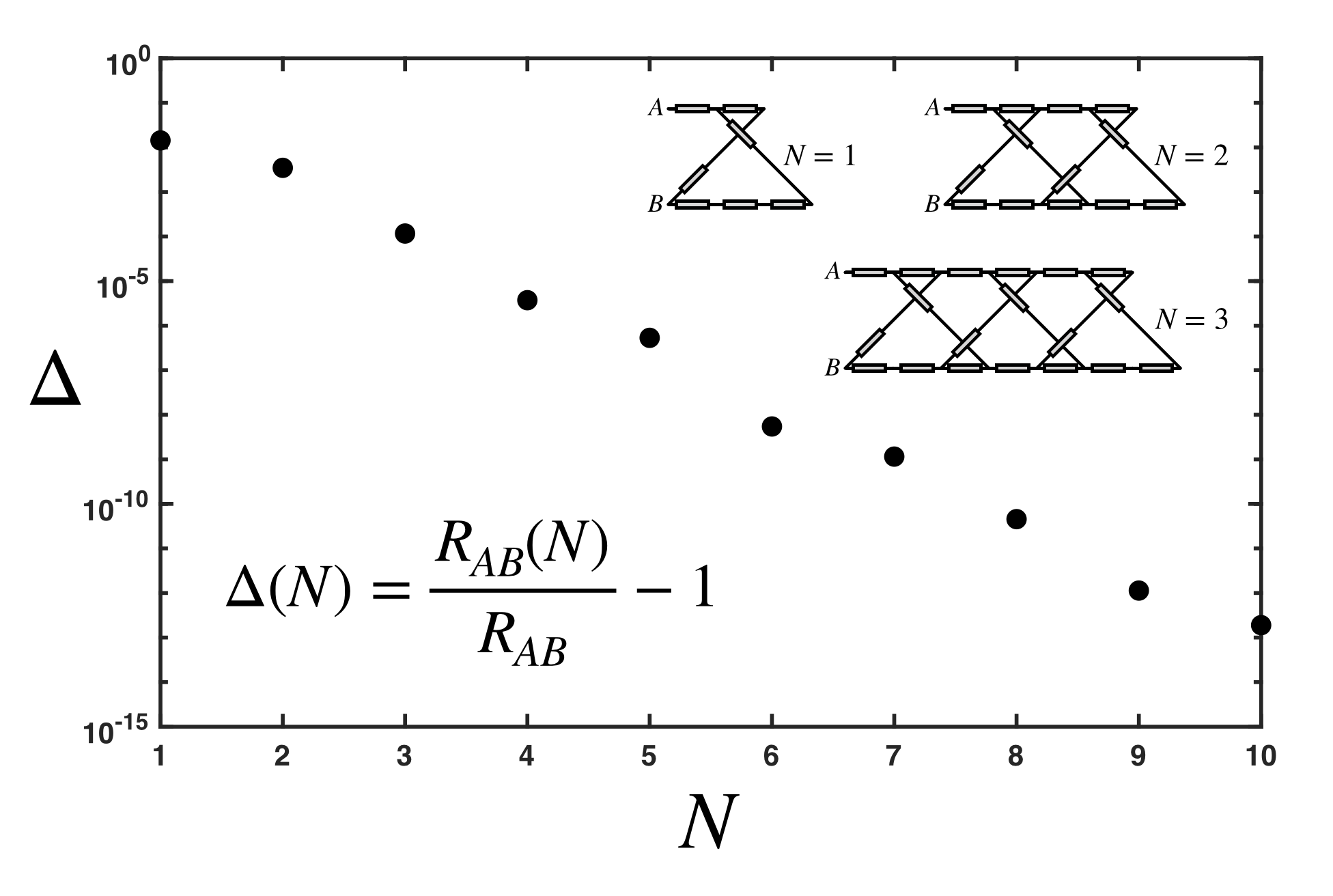}
\caption{We verify the analytical result for $R_{AB}$ with the numerical results for $R_{AB}(N)$ where $N$ is the size of the finite resistive network. The convergence can be quantified by $\Delta = R_{AB}(N)/R_{AB} - 1$, which approaches $0$ exponentially fast.}
\label{confirm_thin}
\end{figure}

To calculate $R_{AB}$ for the asymmetric twisted ladder, we apply the extended $\Delta$-Y transformation mentioned in Section \ref{toolkit} (see Fig. \ref{repeat_thin}A). Assume that the equivalent resistance between nodes $A$, $B$ and $C$ converge in the infinity limit, the trick of adding one more unit cell can be used as shown in Fig. \ref{repeat_thin}B, which gives us a set of three consistency equations:
\begin{equation}
\begin{split}
R_{AB} &= r_\alpha + r_\beta = 1 + r_\beta + 
\frac{(1+r_\alpha)(1+r_\gamma)}{2+r_\alpha+r_\gamma} \ \ , \ \ 
\\
R_{BC} &= r_\beta + r_\gamma = \frac{1+r_\alpha + r_\gamma}{2+r_\alpha+r_\gamma} \ \ , \ \ 
\\
R_{CA} &= r_\gamma + r_\alpha = 1 + r_\beta + \frac{r_\alpha (2+r_\gamma)}{2+r_\alpha +r_\gamma}\ \ . \ \ 
\end{split}
\end{equation}
Three equations for three real positive unknowns $r_\alpha$, $r_\beta$, $r_\gamma$ can be solved analytically:
\begin{equation}
\begin{split}
r_\alpha &= \frac{-1+\sqrt{7} + \sqrt{2\sqrt{7}} }2 \ \ , \ \ 
\\
r_\beta &= \frac{1-\sqrt{7} + \sqrt{2\sqrt{7}} }2 \ \ , \ \ 
\\
r_\gamma &= \frac{-1+\sqrt{-7+4\sqrt{7}}}2 \ \ . \ \ 
\end{split}
\end{equation}
Thus we arrive at
\begin{equation}
R^{(asym)}_{AB}=\sqrt{2\sqrt{7}} \approx 2.30\Omega \ \ , \ \ 
\label{result_thin}
\end{equation}
which is in agreement with \eqref{estimate_thin}. We also confirm this results with numerical evaluation (see Fig. \ref{confirm_thin}). 

\begin{figure}[!htb]
\centering
\includegraphics[width=0.45\textwidth]{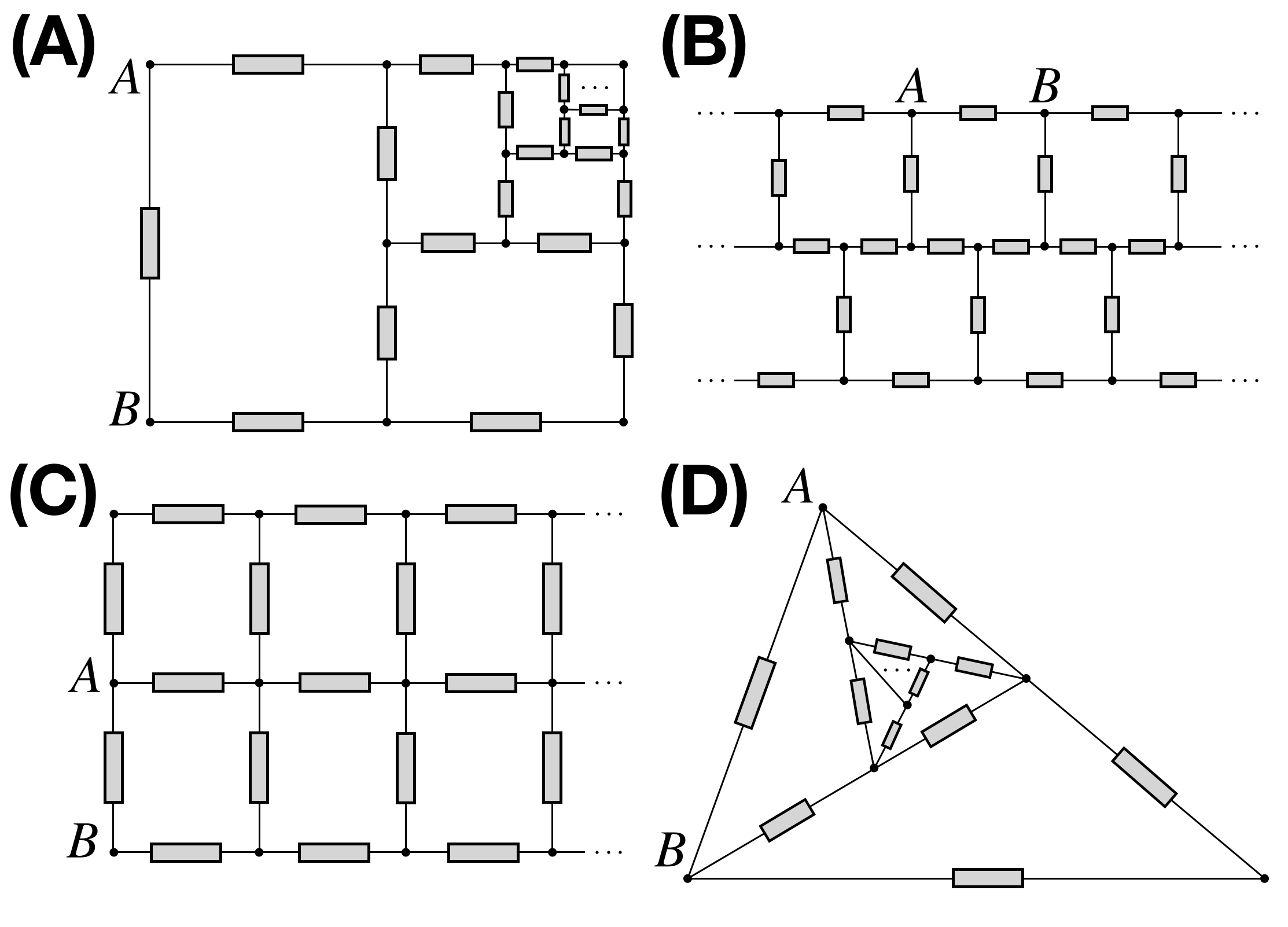}
\caption{All resistors are identically $1\Omega$. (A) Half-and-half network. (B) Brick wall tiling network. (C) 3-parallel ladder network. (D) Triangular spiral fractal network.}
\label{more_thin}
\end{figure}

The method can also be applied to many different topologies, such as those in Fig. \ref{more_thin} which are linear linking with number of nodes $\mathcal{N}=3$ between unit cells. See Table \ref{more_thin_table} for the list of equivalent resistances $R_{AB}$.

\begin{table}[H]
  \begin{center}
    \begin{tabular}{c|c} 
      Circuit & Equivalent Resistance $R_{AB}$ (in $\Omega$) \\
      \hline
      A & $\frac{-3 -\sqrt5 + \sqrt{38 + 14\sqrt5}}4 \approx 0.77$ \\
      B & $1-\sqrt{\frac{7 + 2\sqrt{5}}{145}} \approx 0.72 $ \\
      C & $-1 + \sqrt{\frac{13 + \sqrt{105}}8 } \approx 0.70 $ \\
      D & $\frac{ 1 - \sqrt5 + \sqrt{2 + 2\sqrt5} }2 \approx 0.65 $ \\
    \end{tabular}
  \caption{Results for $R_{AB}$ of infinite resistive circuits given in Fig. \ref{more_thin}.}
  \label{more_thin_table}
  \end{center}
\end{table}

\subsection{$\mathcal{N}=4$ Infinite Symmetric Twisted Ladder \label{section_dense}} 
\begin{figure}[!htb]
\centering
\includegraphics[width=0.45\textwidth]{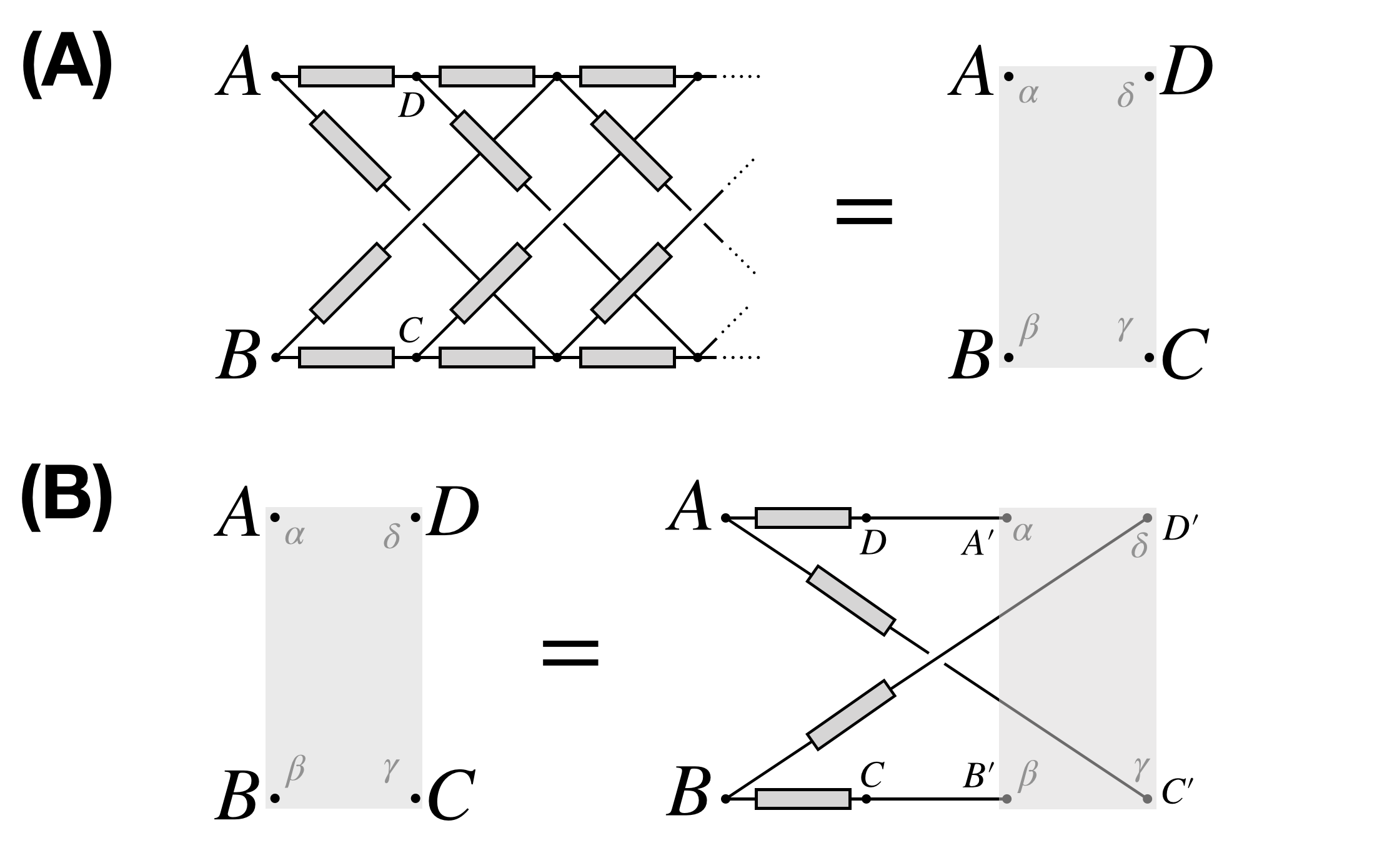}
\caption{(A) The infinite resistive network between nodes A, B, C and D is represented by a ``black box'' with four terminals $\alpha$, $\beta$, $\gamma$ and $\delta$. (B) The trick of adding one more unit cell to the network, where the equal sign is valid under the assumption of convergence.}
\label{repeat_dense}
\end{figure}

We represent the infinite resistive network between nodes A, B, C, and D by a ``black box'' with four terminals $\alpha$, $\beta$, $\gamma$ and $\delta$ (see Fig. \ref{repeat_dense}A). Assume that the equivalent resistances between nodes $A$, $B$, $C$, and $D$ converge in the infinity limit, the trick of adding one more unit cell can be used as shown in Fig. \ref{repeat_dense}B.

\begin{figure}[!htb]
\centering
\includegraphics[width=0.45\textwidth]{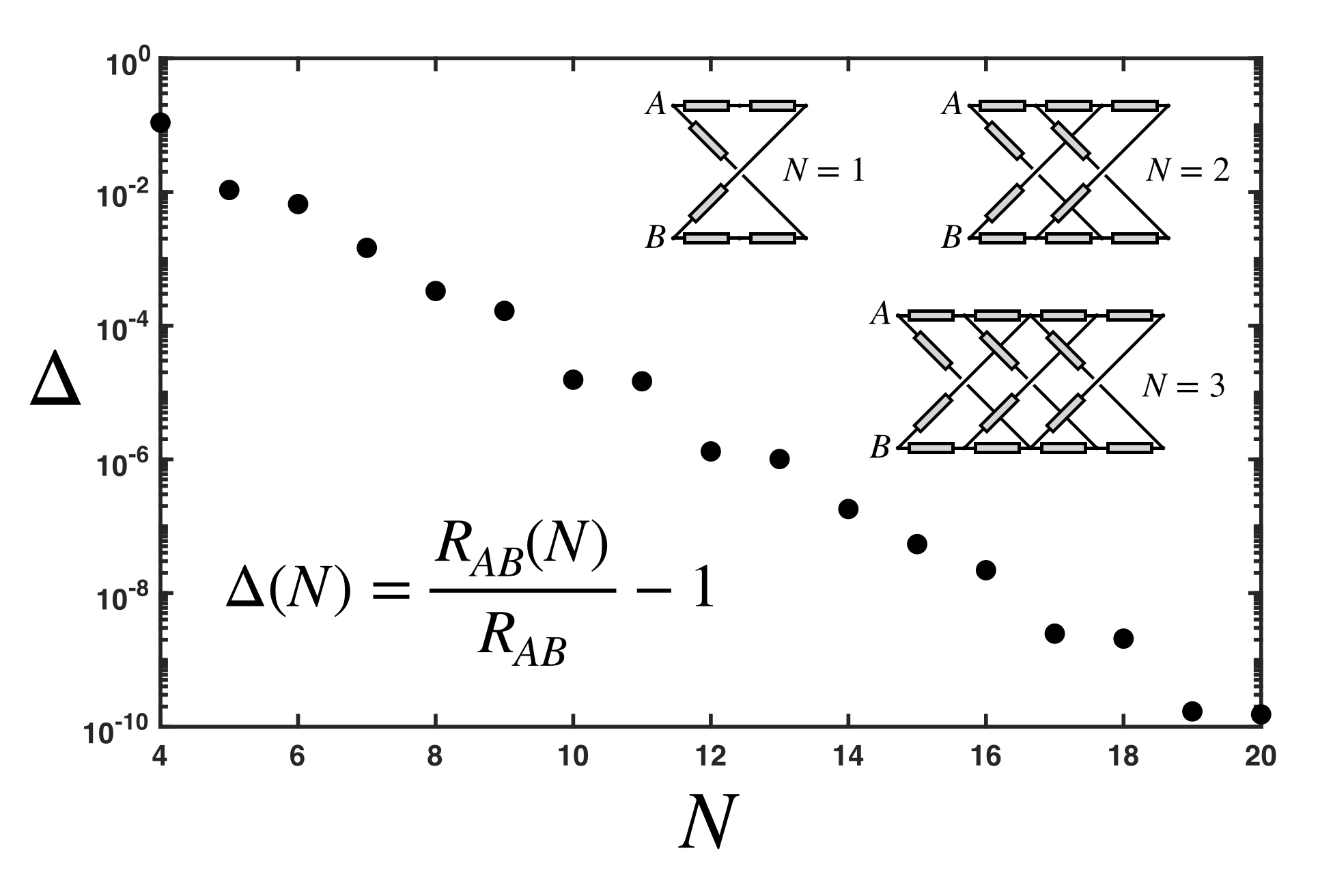}
\caption{We verify the analytical result for $R_{AB}$ with the numerical results for $R_{AB}(N)$ where $N$ is the size of the finite resistive network. The convergence can be quantified by $\Delta = R_{AB}(N)/R_{AB} - 1$, which approaches $0$ exponentially fast.}
\label{confirm_dense}
\end{figure}

\begin{figure}[!htb]
\centering
\includegraphics[width=0.45\textwidth]{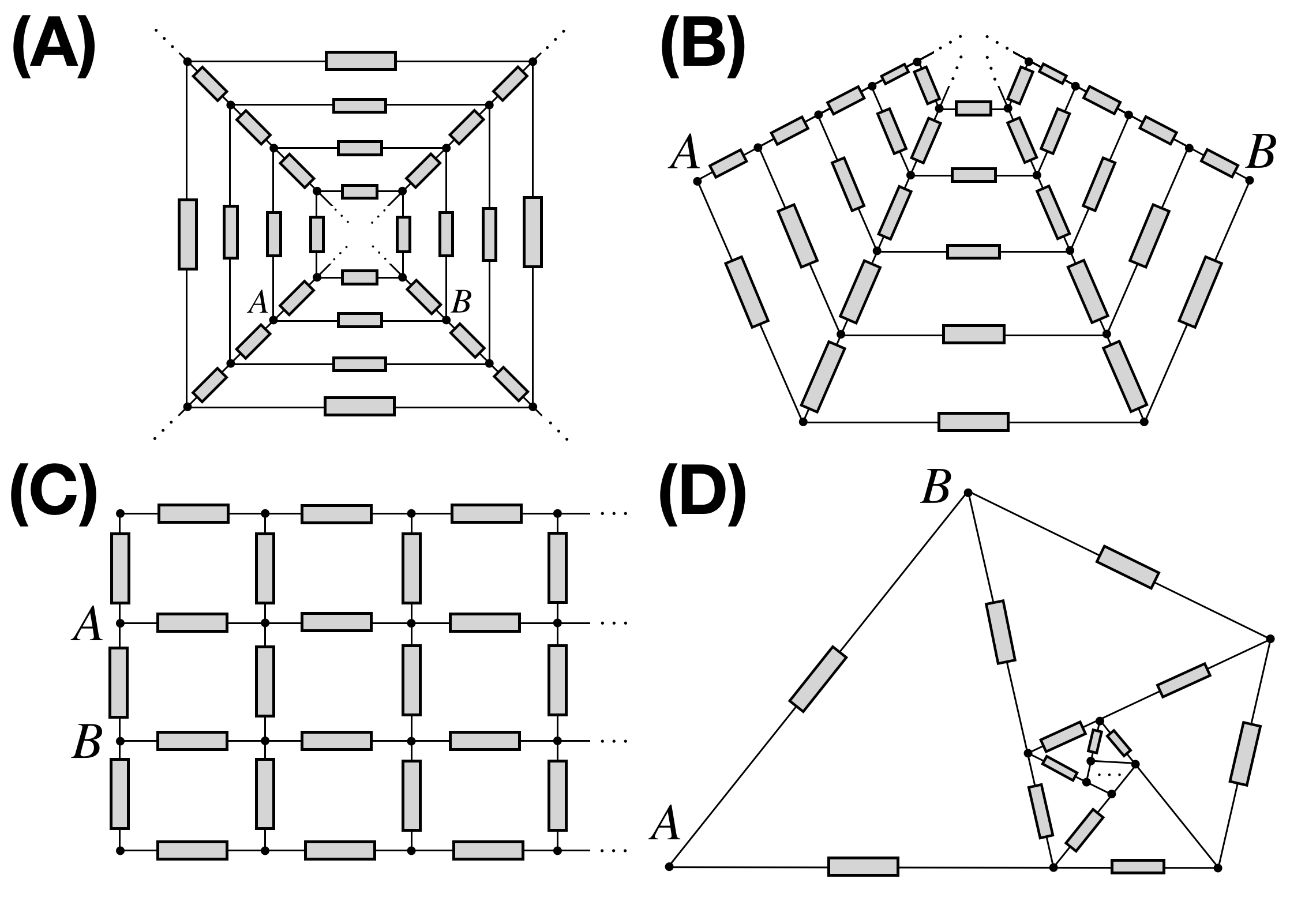}
\caption{All resistors are identically $1\Omega$. (A) Square spider-web network. (B) Pentagonal inception network. (C) 4-parallel ladder network. (D) 4th-order spiral fractal network.}
\label{more_dense}
\end{figure}

Voltage path summation and current conservation are applied on this topology to obtain a set of six independent equations, and the symmetry of the problem reduces the number of variables down to four: $R_{\alpha\beta}$, $R_{\alpha\gamma} = R_{\beta\delta}$, $R_{\alpha\delta} = R_{\beta\gamma}$, and $R_{\gamma\delta}$. Using the toolkit we develop in Section \ref{toolkit}, we can obtain the following set of equations:
\begin{equation}
\begin{split}
    &R_{AB} = R_{\alpha\beta} = \begin{Bmatrix} 1 + G_{\alpha\beta}^{\alpha\gamma} & - (1 + G_{\alpha\beta}^{\delta\beta}) & 1 + G_{\alpha\beta}^{\gamma\beta} \\ 1 + G_{\alpha\delta}^{\alpha\gamma} & 1 + G_{\alpha\delta}^{\beta\gamma} & G_{\alpha\delta}^{\gamma\beta} \\ -(1 + G_{\gamma\beta}^{\gamma\alpha}) & -(1 + G_{\gamma\beta}^{\delta\beta}) & 2 + R_{\gamma\beta}
    \end{Bmatrix} \ \ , \ \ \\
    &R_{CD} = R_{\gamma\delta} = \begin{Bmatrix} G_{\beta\alpha}^{\delta\beta} & G_{\beta\gamma}^{\alpha\gamma} & R_{\alpha\beta} \\ 2 + G_{\delta\alpha}^{\delta\beta} & G_{\delta\alpha}^{\alpha\gamma} & G_{\delta\alpha}^{\beta\alpha} \\ 2 + G_{\delta\gamma}^{\delta\beta} & 2 + G_{\delta\gamma}^{\alpha\gamma} & G_{\delta\gamma}^{\beta\alpha}
    \end{Bmatrix} \ \ , \ \ \\
    &R_{AD} = R_{\alpha\delta} = \begin{Bmatrix} -1 & 0 & 1 \\ 1 + R_{\alpha\gamma} & G_{\alpha\gamma}^{\beta\delta} & 0 \\ 1 + G_{\beta\gamma}^{\alpha\gamma} + G_{\alpha\delta}^{\alpha\gamma} & 2 + G_{\beta\gamma}^{\beta\delta} + G_{\alpha\delta}^{\beta\delta} & 0
    \end{Bmatrix} \ \ , \ \ \\
    &R_{BC} = R_{\beta\gamma} = \begin{Bmatrix} -1 & 0 & 1 \\ 1 + R_{\beta\delta} & G_{\beta\delta}^{\alpha\gamma} & 0 \\ 1 + G_{\alpha\delta}^{\beta\delta} + G_{\beta\gamma}^{\beta\delta} & 2 + G_{\alpha\delta}^{\alpha\gamma} + G_{\beta\gamma}^{\alpha\gamma} & 0
    \end{Bmatrix} \ \ , \ \ \\
    &R_{AC} = R_{\alpha\gamma} = \begin{Bmatrix} 1 + G_{\alpha\beta}^{\alpha\gamma} & G_{\alpha\beta}^{\beta\delta} & G_{\alpha\beta}^{\gamma\beta} \\ - 1 - G_{\beta\gamma}^{\alpha\beta} & -G_{\beta\gamma}^{\beta\delta} & 1 + R_{\beta\gamma} \\ 1 + G_{\alpha\delta}^{\alpha\gamma} & 2 + G_{\gamma\delta}^{\beta\delta} & G_{\alpha\delta}^{\gamma\beta}
    \end{Bmatrix} \ \ , \ \ \\
    &R_{BD} = R_{\alpha\beta} =  \begin{Bmatrix} 1 + G_{\beta\alpha}^{\beta\delta} & G_{\beta\alpha}^{\alpha\gamma} & G_{\beta\alpha}^{\delta\alpha} \\ - 1 - G_{\alpha\delta}^{\beta\alpha} & -G_{\gamma\beta}^{\alpha\gamma} & 1 + R_{\alpha\delta} \\ 1 + G_{\beta\alpha}^{\beta\delta} & 2 + G_{\delta\gamma}^{\alpha\gamma} & G_{\beta\gamma}^{\delta\alpha}
    \end{Bmatrix} \ \ , \ \ \\
\end{split}
\end{equation}
with
\begin{equation}
    \begin{Bmatrix} a_1 & b_1 & c_1 \\ a_2 & b_2 & c_2 \\ a_3 & b_3 & c_3
    \end{Bmatrix} = \frac{\sum^3_{i,j,k=1} \epsilon_{ijk} a_i b_j c_k}{\sum^3_{i,j,k=1} \epsilon_{ijk} a_i b_j} \ \ , \ \  
\end{equation}
where $\epsilon_{ijk}$ is the Levi-Civita symbol. From the set of equations, the real positive $R_{\alpha \beta}$, $R_{\alpha \gamma}$, $R_{\alpha \delta}$, $R_{\gamma \delta}$ values can be found analytically:
\begin{equation}
\begin{split}
&R_{\alpha \beta}=\sqrt{-1 + 2\sqrt2}  \ \ , \ \ 
\\
&R_{\alpha \gamma} = \frac{-4 + 2 \sqrt 2 + \sqrt 5 + \sqrt{5+4\sqrt 2}}{4} \ \ , \ \ 
\\
&R_{\alpha \delta} = \frac{\sqrt 5 -2\sqrt 2 + \sqrt{5+4\sqrt 2}}{4} 
\ \ , \ \ 
\\
&R_{\delta \gamma} = - 1 + \sqrt{4\sqrt2 -2}  \ \ . \ \
\end{split}
\end{equation}
Thus we arrive at
\begin{equation}
R_{AB}=\sqrt{-1 + 2\sqrt{2}} \approx 1.35\Omega \ \ , \ \ 
\label{result_dense}
\end{equation}
which is in agreement with \eqref{estimate_dense}. We also confirm this result with numerical evaluation (see Fig. \ref{confirm_dense}). This same method of using voltage path summation and current conservation to obtain a set of consistency equations can also be applied to many different topologies, such as those in Fig. \ref{more_dense} which are linear linking with number of nodes $\mathcal{N}\geq 4$ between unit cells. See Table \ref{more_dense_table} for the list of equivalent resistances $R_{AB}$.

\begin{table}[!htb]
  \begin{center}
    \begin{tabular}{c|c}
      Circuit & Equivalent Resistance $R_{AB}$ (in $\Omega$) \\
      \hline
      A & $\frac{3\big(-19 + 18\sqrt{3} + 3\sqrt{122 + 65\sqrt{3}} \big)}{376} \approx 0.48$ \\
      B & $\frac{-2+\sqrt{23+\sqrt{17}}}2 \approx 1.60$ \\
      C & $\frac{-2+\sqrt{7+\sqrt{17}}}2 \approx 0.68$ \\
      D & $R_{AB} = \frac{2-x^2 }{(1-x)x} \approx 0.67$, \\ & where $x \approx 1.65$ satisfies \\ & 
         $x^8 - 4x^7 + 4x^6 + 9 x^5 - 27 x^4$ \\ & $+ 18 x^3 + 16 x^2 - 32 x + 16 = 0$ \\ 
    \end{tabular}
  \caption{Results for $R_{AB}$ of infinite resistive circuits given in Fig. \ref{more_dense}.}
  \label{more_dense_table}
  \end{center}
\end{table}

It should be noted that for a highly entangled topology linking linearly, while it is still possible to use the resistive relations to calculate the equivalent resistance, there will be $\mathcal{N}(\mathcal{N}-1)/2$ consistency equations with $\mathcal{N}(\mathcal{N}-1)/2$ in general thus the algebraic manipulation gets messy quickly as $\mathcal{N}$ increases. Even if all resistors are identically $1\Omega$, the solution might not have any algebraic representation \cite{Abel_Impossibility} (for example, $R_{AB}$ of the network given in Fig. \ref{more_dense}D), though it should still be an algebraic number \cite{Sturmfels_Polynomials}.

We should also mention that the equivalent resistances of linearly linking infinite resistive networks can be calculated with a linear algebra method via transmission matrix \cite{Matthaei_Network}, which will be discussed elsewhere \cite{Infinite_AC_Twist}. This approach, however, is problematic and even inapplicable for looping topologies such as the 4th-order spiral fractal network (see Fig. \ref{more_dense}D).

\section{Discussion}
The potential of the method discussed above expands to infinite network consisting of inductors and capacitors, yielding a physical explanation for the convergence of $Z$ in the existence of infinitesimal resistances as well as the analytical expression of the point of convergence. It can also be utilized to provide a rigorous explanation for Rayleigh's theorem for system with only resistors, and to expand this well-known theorem to system with both inductors and capacitors. Those applications will be discussed elsewhere \cite{Infinite_AC_Twist}.

\begin{acknowledgments}

We thank Tuan K. Do, Hiep T. Vu, Duy V. Nguyen, and xPhO journal club for their valuable feedback and insight throughout every stage of this research.

\end{acknowledgments}

\bibliography{main}
\bibliographystyle{apsrev4-2}

\end{document}